\documentstyle[11pt,aaspp4]{article}

\def\gtrapprox{\;\lower 0.5ex\hbox{$\buildrel >
    \over \sim\ $}}             
\def\lessapprox{\;\lower 0.5ex\hbox{$\buildrel < \over \sim\ $}}

\begin{document}

\title{Search for Free-Free Absorption Cutoffs from Tori in Three
Type 2 Active Galactic Nuclei}

\author{Richard Barvainis \& Colin Lonsdale}
\affil{MIT Haystack Observatory, Westford, MA 01886 USA\footnote
{Radio Astronomy at the Haystack Observatory of the 
Northeast Radio Observatory Corporation (NEROC) is supported by a grant 
from the National Science Foundation}}

\centerline{Email: reb@dopey.haystack.edu, cjl@dopey.haystack.edy}
\begin{abstract}
We have observed three Type 2 radio loud active galactic nuclei using
the VLBA at five frequencies between 1.6 GHz and 15 GHz, in a search for
the free-free absorption cutoff predicted by some models for
optically-obscuring dusty molecular tori.  We find no evidence for sharp
cutoffs toward low frequencies in any of the sources.  However, one
source, OW637, has VLBI components which are steeply falling toward high
frequencies ($S_{\nu} \propto \nu^{-2}$).  
\end{abstract}

\keywords{Galaxies: Active} 

\section{Introduction}

By now the standard dusty molecular torus scenario for unifying Type 1
(broad line) and Type 2 (narrow line) active galactic nuclei via
obscuration/orientation effects will be familiar to most readers
(otherwise, see Antonucci 1993 for a comprehensive review).  The physics
of the hypothesized torus was modeled in detail in a series of papers by
Krolik and collaborators (Krolik and Begelman 1986, 1988; Krolik and
Lepp 1989, hereafter KL).  Their general picture is that of a torus
composed of small, dusty molecular clouds lying about a parsec from the
central engine.  They are unusual molecular clouds, according to KL,
having electron fractions $\sim 10^{-3}$ and temperatures $\sim 10^3$ K.

To date the torus model has not been satisfactorily tested.  Barvainis
and Antonucci (1994) searched unsuccessfully for CO absorption from a 
torus against the compact radio nucleus of the Type 2 radio 
galaxy Cygnus A.  The result, while unexpected at the time, was  
subsequently rationalized within the torus framework by Maloney, 
Begelman, \& Rees (1994).

One testable prediction from torus models is that the torus should
free-free absorb radio photons below a certain frequency as a result of
the relatively high electron densities -- provided, of course, that the
central radio source is compact enough to be blocked by the torus.  The
inner edge of the torus is at a radius of about 1 pc, for a central UV
luminosity of $10^{44}$ erg s$^{-1}$ (scaling as $L^{1/2}$).  The height
of the torus material goes roughly as $h \approx 0.7r$.  According to
the KL model, the cutoff frequency is about $5-10$ GHz.  However,
Neufeld, Maloney, \& Conger (1994, hereafter NMC) found a much lower
electron fraction in the torus relative to that of KL ($10^{-5}$ versus
$10^{-3}$), with the discrepancy arising because KL underestimated the
destruction rate of the molecular ion H$_3^+$.  NMC predict instead a
warm atomic region lying near the inner edge of the torus, having an
electron fraction $\sim 10^{-2}$ (see also Maloney 1996).
Coincidentally, however, the resulting free-free absorption is the same
as for KL:  the cutoff frequency is predicted to be in the few to 10 GHz
range.

In a search for this predicted cutoff, we observed the radio cores
of three narrow line radio galaxies at a range of frequencies between
1.6 GHz and 15 GHz, using the NRAO\footnote{The National Raio Astronomy
Observatory is operated by Associated Universities Inc., under
cooperative agreement with the National Science Foundation.}  Very Long
Baseline Array (VLBA).  We report the results below.

\section{Source Selection and Observations}

The ideal source for this search would be a Type 2 AGN, i.e., one in 
which the nucleus (central optical continuum source, broad line region) 
is obscured by the torus.  It would be reasonably strong in the radio regime, and
sufficiently compact to have significant flux on VLBI scales.  To further 
select for compactness the radio spectrum would be flat or inverted, 
indicating a partially self-absorbed synchrotron source.

Mrk 348 (NGC 262; $z=0.014$) was chosen as our top-rated object, for its
strong, inverted VLBI core source (Neff \& de Bruyn 1983), and its
nearness (providing good linear resolution).  It has a Seyfert 2 optical
spectrum in unpolarized light, with clear broad emission lines in
polarized (reflected) light, indicating the presence of a hidden Seyfert
1 nucleus (Miller \& Goodrich 1990).  Additional evidence for the
presence of a dusty torus has been offered by Simpson et al (1996).  Its
host galaxy is an early-type spiral.

Two other sources were selected for observation because of their
availability during the observing time allocated for Mrk 348.  These are
OW637 and 3C99, both of which are narrow-line radio galaxies.  OW637
($z=0.227$) has a strong, flat-spectrum core which dominates the radio
emission.  3C99 ($z = 0.426$) is a lobe-dominated radio galaxy with very
asymmetric lobe fluxes, having a VLBI core at 1.6 GHz with peak flux
density 60 mJy.  The core is steep spectrum, making this source less
than ideal for observation of very compact structure likely to be
contained within a torus.

The observations were conducted during a 12.5 hour period
on April 5/6, 1995, under VLBA
project
code BB21.  The usefulness of the results in this project depended
heavily on image fidelity, and an observing strategy was chosen which
optimized the $(u,v)$ coverage.  We used the frequency agility of the VLBA to
switch rapidly between the five observing frequencies:  1630 MHz, 2270
MHz, 4990 MHz, 8420 MHz and 15360 MHz.  The observing frequency was
switched typically every 2 minutes, and combined with interleaving of
target sources in time, all 15 source/frequency combinations received
good $(u,v)$ coverage, with no major holes.  In addition to the program
sources, 3C84 and 1739+522 were observed as fringe-finders.

The data were correlated in Socorro, and were calibrated in the normal manner
using the VLBA log data.  The data were edited, fringe-fitted and imaged in 
AIPS with no significant problems, except that 3C99 at 15 GHz was too weak for
fringe-fitting and had to be abandoned.  The RMS noise levels on the images
are generally in the range 0.2-0.5 mJy/beam, except for OW637 for which dynamic
range limits of 200-500 result in rms fluctuations at the 3 mJy/beam level.
Component flux densities are reported in Table 1.   At 8.4 GHz and 15 GHz
the fluxes are taken from maps convolved to the same beamsize as the
5 GHz maps, in order to avoid resolution effects and facilitate measurement 
of spectral indicies.

We obtained additional X-band data on OW637 from the United States
Naval Observatory (USNO) Radio Reference
Frame Image Database program
(courtesy Alan Fey).
These observations are of longer duration and 
higher quality than our own X-band measurements (contour image 
can be seen in Fey et al 1996).  The
observing date was 1994 July 8.  Four separate observing frequencies
within the X-band were used.  See Table 1 for details.

\section{Results and Discussion}

\subsection{Low Frequency Cutoffs}

The angular resolutions differ by a factor of order 10 from L-band to U-band,
limiting the range of component sizes and separations for which reliable
5-point spectra can be derived.  Only compact, well-separated components are
analyzed below.  While high-frequency spectral information at higher resolution
exists in our dataset, no dramatic spectral gradients were evident, and no 
detailed analysis was performed.

3C99 shows double structure in the VLBI core which is resolvable at all
frequencies, with a separation of 14 mas.  A weak third component is also
visible in some of the maps of 3C99.  OW637 has four VLBI components
(see below).  Mrk 348 was resolved into two components only in the
highest resolution map (15 GHz), with a component separation of 1
mas.  For 3C99 and OW637 we plot radio continuum spectra of all
components that could be well-resolved (Figures 2 and 3), while for Mrk
348 the components are not separable at the lower frequencies and we
plot only the total flux (Figure 1).

None of the spectra show the sharp low-frequency cutoff expected for
free-free absorption.  Component B in OW637 rolls over from 5 GHz to 2.3
GHz, but the spectral index of $+1$ is not high enough to eliminate
other possibilities such as synchrotron self-absorption (it is worth
bearing in mind, however, that this represents a lower limit to the spectral
index because of the finite sampling of the spectrum).  Therefore we
have not found any new evidence for the existence of a torus.  The
possibilities for explaining this result are:  
\begin{itemize} \item The
torus is not present in these objects.  Given the growing body of direct
evidence for anisotropic obscuration, combined with persuasive arguments
on which orientation-related unified schemes are based, it seems
probable that something akin to a torus does in fact exist, and we
regard this possibility as unlikely.  
\item The torus is there, but its
parameters are such that the cutoff occurs below our lowest observing
frequency of 1.6 GHz,
or above our highest observing frequency of 15 GHz (so that the
true core is never seen).  It may be 
worthwhile to investigate possible torus conditions which would give rise 
to significantly lower (or higher) free electron densities than the models 
of KL or NMC.  
If the inner edge of the torus resides for some reason outward of the
$\sim 1$ pc inner radius of these models, the free-free optical depth will
be lower because of a reduction in radiation pressure (upon which $\tau_{\rm
f-f}$ depends linearly).  Maloney (1996) invoked this effect as a
possible explanation for the absence of free-free absorption at
gigahertz frequencies in Cygnus A (see that paper for expressions
relating $\tau_{\rm f-f}$ and various physical parameters of the torus).
\item The test is not applicable because the bulk of the radio emission
lies outside the torus and therefore cannot be absorbed by it, or the 
torus orientation is not ``side-on''.  The former could occur if the jet
material we observe lies more than a parsec or so from the true central
engine.  This seems plausible for 3C99, which has no strong
self-absorbed component identifiable as a ``core", but less so for Mrk
348 and OW637, which do.  The ``core" component would then have to be a
standing shock or similar feature in the flow some distance downstream
from the center.  Hints of such a phenomenon may have been seen in other
sources (Marscher 1997).  Orientation may be more of an issue for Mrk
348 and OW637, despite our optical selection for torus obscuration of
the inner regions.  Both objects have strong, variable (see Neff \& De
Bruyn 1983 for Mrk 348, and below for OW637) cores, and weak or
non-existent lobes.  These properties suggest that our line-of-sight may
be near the axis of a relativistic jet and therefore, in the unified
picture, perpendicular to the plane of the torus.  Future experiments,
perhaps using phase referencing to enhance sensitivity, might profitably
target weak, quiescent cores of lobe-dominated narrow line radio galaxies.
\end{itemize}

Free-free absorption on small scales has been discussed recently for
several radio galaxies.  Probably the most convincing case known to us
is that of Centaurus A, where Jones et al (1996) found that the spectrum
of the milliarcsecond core is probably highly inverted, $\alpha \sim +4$
($S_{\nu} \propto \nu^{\alpha}$) between 2.3 GHz and 8.4 GHz, with the
main uncertainty being a possible misregistration of the images at the
two frequencies.  Another good case is that of NGC 1275 (3C84), where
the northern jet (the counterjet) is resolved, yet has an inverted
spectrum (Vermeulen, Readhead, \& Backer 1994; Walker, Romney, \& Benson
1994; Levinson, Laor, \& Vermeulen 1995).  The inferred size of the
free-free absorbing screen, or torus, is less than a parsec for Cen A,
and a few parsecs for NGC 1275.  Hydra A and NGC 4261 may also possess
components that are free-free absorbed (Taylor 1996; Jones \& Wehrle
1997).

Our strategy was to look for the most direct evidence of free-free
absorption, that being a spectrum well-sampled over a range of
frequencies showing a clear and sharp cutoff toward low frequencies.
Such a cutoff was not seen in the three Type 2 AGNs observed.  The most
probable explanation is that the emission measure ($\int n_e^2 dL$) 
through the torus is lower
than current models suggest, at least in these objects, though it is
possible that our observations failed to probe the torus due to
geometrical considerations.  

\subsection{Unusually steep components in OW637}

The maps of OW637 show several components, arranged roughly along a line
(four components are resolvable at C- and X-bands; see Figure 4).
Components A, B, and D (same convention as used by Bartel et al 1984)
appear to be unresolved at all frequencies.  Component C is extended at
C-, X-, and U-bands, making its spectrum difficult to determine (we do
not report flux values for component C for this reason).  Therefore we
only show spectra of components A, B, and D in Figure 3.  Component D is
relatively flat over the limited frequency range where it is not blended
with component C.  However, components A and B both have spectra that
are unusually steep (i.e., steeply falling toward high frequencies), at
least over part of the spectral range.  Component D appears to have
varied since the measurements of Bartel et al (1984).  We find a flux
density of $1950\pm 100$ mJy at 8.4 GHz, compared with their $850\pm 70$
mJy at 8.3 GHz.

Component A appears to be steep throughout the measured range, with the
segment between C- and X-bands having spectral index $\alpha_{\rm CX} =
-2.05\pm 0.14$.  The steep slope is apparent even within the
closely-spaced X-band measurements (see inset to Figure 3).  Component B
has a spectral peak at 5 GHz and steepens rapidly toward high
frequencies, with $\alpha_{\rm XU} = -2.22\pm 0.11$ between 8.4 and 15.4
GHz.  Such steeply falling spectra appear to be quite unusual.  Among
the 518 total flux density (i.e., large-beam) spectra of extragalactic
radio sources assembled by K\"uhr et al (1981), none have spectra
approaching this steepness.  However, summing all the components in OW637
will produce a fairly flat spectrum, so it is possible that very steep
components like A and B may only be observable with VLBI.

Typical spectral indices for {\it extended} structures in extragalactic
radio sources are in the range $-1.3 \lesssim \alpha \lesssim -0.5$,
with the steepest observed index being about $-2$ (Kellerman \& Owen
1988).  For a sample of ultrasteep radio sources compiled by
R\"ottgering et al (1994), the steepest have spectral indices $\sim
-1.6$.  Steep spectra in diffuse sources are generally thought to be due
to synchrotron aging of electrons that have drifted far from any source
of reacceleration.  Spectra of VLBI components typically run from
inverted ($\alpha \sim +1$, due to synchrotron self-absorption and
component superposition), to moderately steep ($\alpha \sim -1$).  We
know of no VLBI components previously reported with slopes as steep as
$-2$.  Synchrotron lifetimes for electrons in luminous, compact
components like those seen in OW637 can be of order years.  However,
current thinking holds that VLBI components are the sites of shocks in
jets, with attendant continuous reacceleration of particles.  It is
conceivable that in OW637 a shock has suddenly turned off, allowing the
decay of electron energies through synchroton losses.  However, this
would have to have happened at nearly the same time in components A
and B to produce the observed steep spectra in both, and therefore seems
unlikely.

A nonrelativistic, adiabatic, strong shock will have a compression
factor $3 \lesssim r \lesssim 4$, and electrons accelerated by the Fermi
mechanism in such a shock will have an electron energy power law index
$-2.5 < s < -2$ ($N(E) \propto E^{s}$, with $s = -{r+2\over r-1}$) (see
Begelman, Blandford, \& Rees 1984).  This produces the canonical range
$-0.75 < \alpha < -0.5$ for the radiation spectral index via $\alpha =
{s+1\over 2}$.  The case of $\alpha \lesssim -2$ found for OW637
requires $s = -5$, and $r = 1.75$.  We are not aware of any specific
acceleration models that predict numbers like these.

Another possibility for producing a steep spectrum is synchrotron
emission from a relativistic Maxwellian electron distribution at
frequencies well above $\nu_T$, the characteristic synchrotron frequency
for electrons of energy $Tm_ec^2$.  At $\nu / \nu_T \sim 100$, the
spectrum is steep and slightly curved (convex), with $\alpha \sim -2$
(using equation 6 of Jones \& Hardee 1979).  This could account for the
shape of component A, but not B, which is too sharply peaked to be fit
by a relativistic Maxwellian.

Pulsars produce radio spectra that are very steep, ($-1 < \alpha < -4$)
via coherent radiation, but quasar radio emission is thought to be
incoherent synchrotron and differs observationally from that of pulsars
in several ways (e.g., much lower levels of polarization, lower
brightness temperatures, normally much flatter spectra).  However, owing
to its two steep-spectrum components OW637 is an unusual case among
quasars and should be looked at more thoroughly.  The present
observations were set up only for total intensity measurements; we plan
to measure the polarization properties and extend the spectral coverage
of OW637 in a future experiment.

\acknowledgments 
We thank Barry Clarke for scheduling observing time for this project on 
the VLBA, and Alan Fey for providing data on OW637.  Steve Reynolds and
Phil Maloney provided valuable advice.

\vglue 1.0truecm
\centerline {\bf FIGURE CAPTIONS}

\figcaption{Continuum spectrum of the total core flux of Mrk 348.  The 15 
GHz map shows a barely resolved double source, which was unresolved at 
all lower frequencies.  \label{fig1}}

\figcaption{Continuum spectra of the two main compact components observed
in 3C99. \label{fig2}}

\figcaption{Continuum spectra of three compact components observed
in OW637. Inset shows USNO Reference Frame data for X-band, and the steep spectrum
evident even over a narrow range of frequencies. \label{fig3}}

\figcaption{5 GHz map of OW637, showing the four components discussed in 
the text.  Contours start
at 10 mJy and increase by factors of 2.
\label{fig4}}

\begin{deluxetable}{lcccccc}
\tablecaption{Component Flux Densities (mJy) \label{tbl-1}}
\tablehead{\colhead{Source} 
& \colhead{1.63 GHz}
& \colhead{2.27 GHz}
& \colhead{4.99 GHz}
& \colhead{8.42 GHz}
& \colhead{15.4 GHz}
& \colhead{Component}}
\startdata
Mrk 348 &$40.7\pm 2.3$&$61.7\pm 4.6$&$115\pm6$&$163\pm 8$&$169\pm 9$ & \nl 
3C99    &$51.4\pm 5.8$&$45.2\pm 3.0$&$30.6\pm1.6$&$21.7\pm 1.7$&... &  \nl 
        &$115\pm 9$&$80.8\pm 4.6$&$29.6\pm1.6$&$ 6.8\pm 1.5$&... &  \nl 
OW637   &...\tablenotemark{a}&$360 \pm 40 $&$117\pm 7 $&$40 \pm 1.5$
\tablenotemark{b}&$< 8$& A \nl 
        &...\tablenotemark {a}&$420 \pm 40 $&$925 \pm47 $&$750 \pm 30$&$196 \pm 10$& B \nl 
        &...\tablenotemark {a}&... \tablenotemark{a}&$1400 \pm70 $&$1950 \pm 100$
	&$1500\pm 70$& D \nl 
\enddata
\tablecomments{Flux uncertainties assume 5\% calibration uncertainty 
added in quadrature to the statistical uncertainty.  Values represent
total component flux densities as reported by AIPS task JMFIT.}
\tablenotetext{a}{Components blended.}
\tablenotetext{b}{Mean value of observations at four frequencies, obtained
from the USNO Radio Reference Frame Image Database.   The frequencies are 
8150 MHz, 8230 MHz, 8410 MHz, and 8550 MHz, and the respective 
fluxes are $41.2\pm 1.5$ mJy, $42.8\pm 1.5$ mJy, $39.6\pm 1.5$ mJy, 
and $35.8\pm 1.5$ mJy.}
\end{deluxetable}

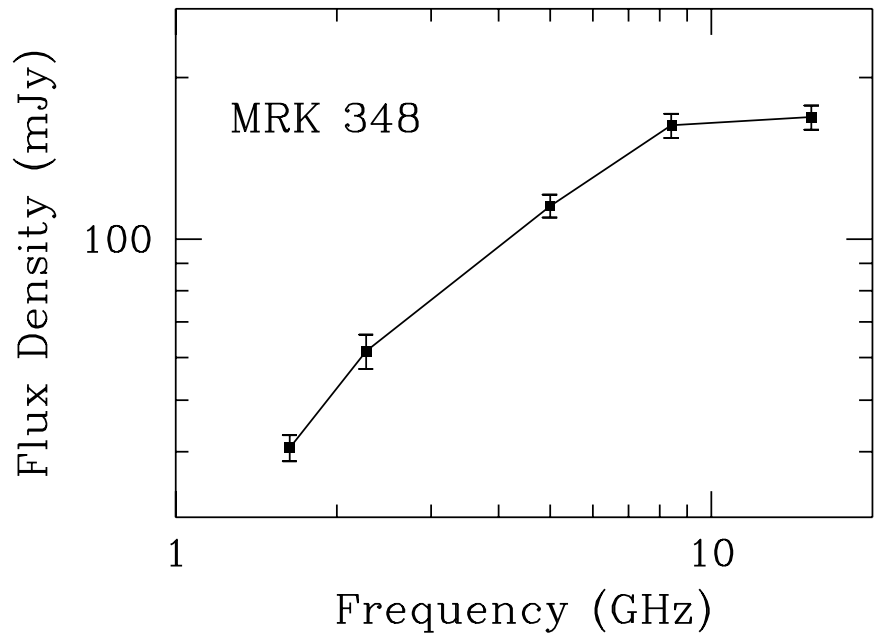
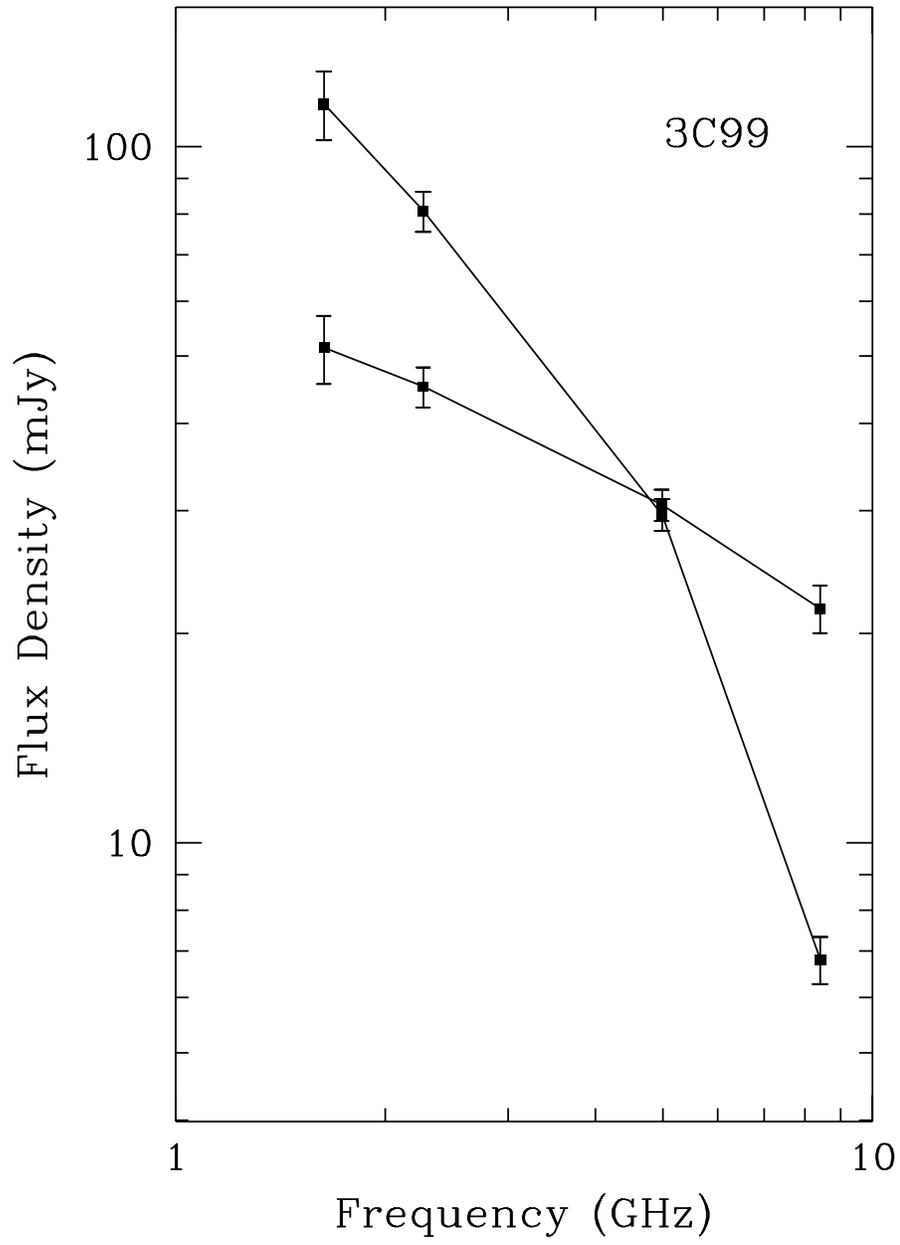
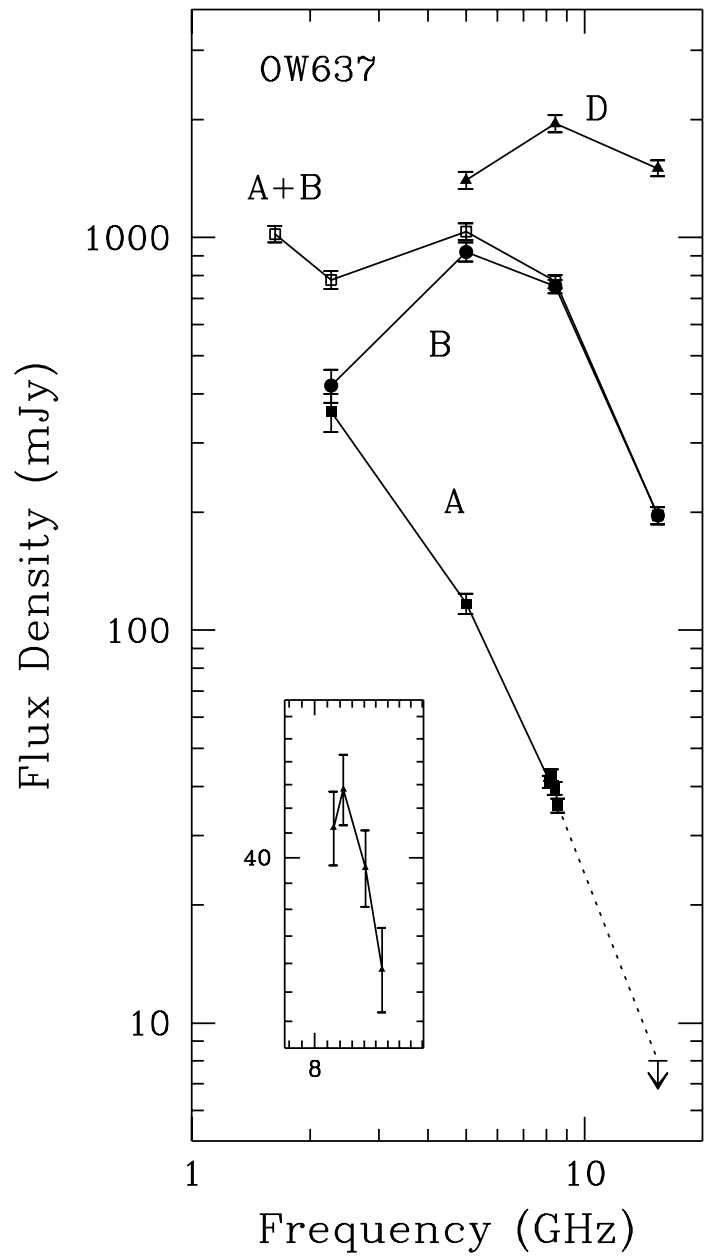
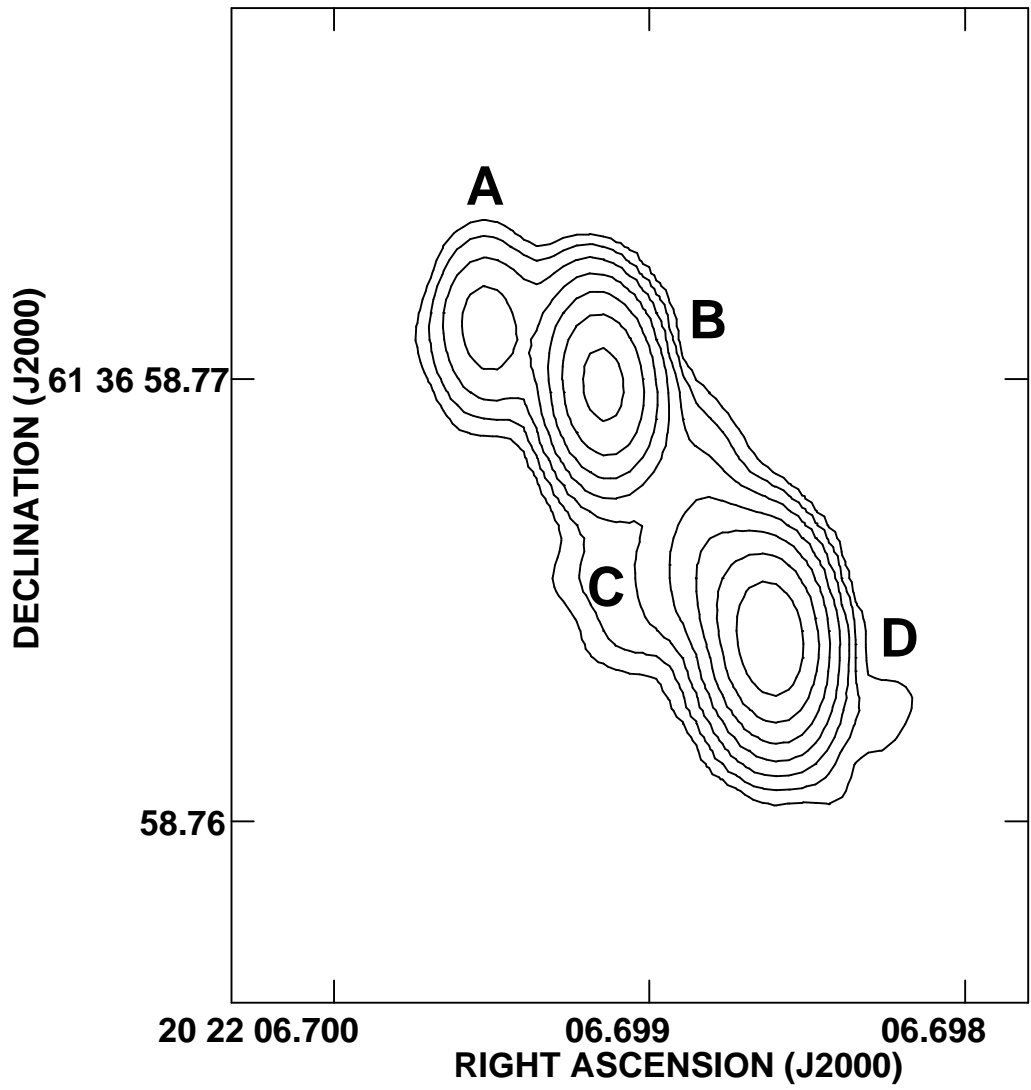
\end{document}